\documentclass[journal]{IEEEtran}
\usepackage{graphicx}
\usepackage{epstopdf}
\usepackage[cmex10]{amsmath}
\usepackage{cite}
\usepackage{amsfonts}
\usepackage[tight,footnotesize]{subfigure}
\usepackage{algorithm}
\usepackage{rotating}
\usepackage{array}
\usepackage{multirow}
\usepackage{algpseudocode}
\usepackage{bm}
\usepackage{amssymb}
\usepackage[bookmarks=false,draft]{hyperref}

\newcolumntype{K}[1]{>{\centering\arraybackslash}p{#1}}
\begin{document}
\title{Hierarchical Online Intrusion Detection for SCADA Networks}
\author{Hongrui~Wang, Tao~Lu, Xiaodai~Dong, Peixue~Li and Michael~Xie\thanks{Hongrui~Wang, Tao~Lu and Xiaodai~Dong are with the Department of Electrical and Computer Engineering, University of Victoria, Victoria, BC, V8P 5C2, Canada, e-mail: hongrui@uvic.ca; taolu@ece.uvic.ca (T.L.); xdong@ece.uvic.ca (X.D.).} \thanks{Peixue~Li and Michael~Xie are with Fortinet Corp., 899 Kifer Road Sunnyvale, CA 94086, USA.}}
\markboth{}%
{Wang \MakeLowercase{\textit{et al.}}: Hierarchical Online Intrusion Detection for SCADA Networks}
\maketitle

\begin{abstract}
We propose a novel hierarchical online intrusion detection system (HOIDS) for supervisory control and data acquisition (SCADA) networks based on machine learning algorithms. By utilizing the server-client topology while keeping clients  distributed for global protection, high detection rate is achieved with minimum network impact. We implement accurate models of normal-abnormal binary detection and multi-attack identification based on logistic regression and quasi-Newton optimization algorithm using the Broyden-Fletcher-Goldfarb-Shanno approach. The detection system is capable of accelerating detection by information gain based feature selection or principle component analysis based dimension reduction.
By evaluating our system using the KDD99 dataset and the industrial control system dataset, we demonstrate that HOIDS is highly scalable, efficient and cost effective for securing SCADA infrastructures.
\end{abstract}


\section{Introduction\label{introduction}}

\IEEEPARstart{H}{istorically}, industrial control systems (ICSs) are proprietary, making massive attacks against them virtually impossibile~\cite{knapp2014industrial}. The standardization of these systems and increasing adoptions of common communication protocols such as TCP/IP, Modbus and DNP3, and common operating systems such as Windows and LINUX, despite of obvious advantages on performance improvement and cost reduction, have made the was-secrative information easier to access. Consequently, modern ICSs are more vulnerable to attacks~\cite{radvanovsky2013handbook}. This is of particular concern to the supervisory control and data acquisition (SCADA) system~\cite{nicholson2012scada}, one of the most important ICSs that typically includes large-scale processes in multiple sites to control critical infrastructures such as smart grid, oil and gas pipelines, water and sewage systems, etc. Breaches of such systems lead to disastrous consequences. The interconnections between the SCADA and external networks such as the Internet and enterprise networks further expose SCADA to large-scale cyber security threats. One of the typical examples is the Stuxnet worm~\cite{langner2011stuxnet}  discovered in 2010, which infected more than $100,000$ computers worldwide and damaged almost one-fifth of nuclear centrifuges in Iran by exploiting four zero-day vulnerabilities of Windows systems.

To protect conventional information technology (IT) networks from cyber attacks, implementing intrusion detection systems (IDSs) has been a common practice~\cite{axelsson2000intrusion}. IDSs are usually signature-based and/or statistical anomaly-based. Signature-based techniques are highly efficient as they identify the malicious data flow through whitelists or rules generated from the characteristics of known attacks while anomaly-based IDSs spot intrusions by comparing data flow with a principle generated by statistic algorithms~\cite{patcha2007overview}. Anomaly-based IDSs are comparatively less efficient as they adopt complex algorithms and consume substantial computing resources. However, a major advantage over signature-based IDSs is that they can detect unknown attacks or mutants of known attacks.


Existing intrusion detection modules for conventional IT networks cannot be directly exploited in SCADA networks due to different network characteristics and system requirements~\cite{zhu2010scada}. For example, SCADA systems emphasize real-time requirements and many SCADA devices have limited computing abilities.
The growing awareness of SCADA security has motivated researches on SCADA-specific IDS. Among them, many SCADA IDSs are signature-based to accomodate the strict real-time constraint and often less computationally powerful devices in the networks~\cite{cheung2007using,oman2008intrusion,carcano2011multidimensional,ten2011anomaly,goldenberg2013accurate,yang2014multiattribute}. On the other hand, many attacks are either unknown in prior or mutants of their original form. Under these circumstances, anomaly-based IDSs, particularly using machine learning algorithms are advantageous.
Yang \textit{et~al.}~\cite{yang2006anomaly} applied an autoassociative kernel regression model along with the statistical probability ratio test and demonstrated the effectiveness of their design on a simulated SCADA system for intrusion detection. Their training dataset consisted of $1,000$ observations, the network traffic statistics of which are from Simple Network Management Protocol. Linda \textit{et~al.}~\cite{linda2009neural} proposed an IDS based on two neural network learning algorithms: the Error-Back Propagation and Levenberg-Marquardt, and tested their model using datasets generated by software tools such as Nmap, Nessus and Metasploit. Valdes \textit{et~al.}~\cite{valdes2009communication} investigated two anomaly detection techniques: pattern-based detection for communication patterns among hosts, and flow-based detection for individual traffic flows, and showed that their methods are capable of identifying basic attacks against the Modbus servers in their distributed control systems testbed. Zhang \textit{et~al.}~\cite{zhang2011distributed} exploited the support vector machine (SVM) technique and artificial immune system tested on the NSL-KDD dataset to evaluate their distributed intrusion detection system for the multi-layer network architecture of smart grid and related SCADA systems. Maglaras \textit{et~al.}~\cite{maglaras2014intrusion} presented their SCADA intrusion detection module based on One-Class SVM, training the network data offline with a dataset of $1,570$ packets. Brushi \textit{et~al.}~\cite{rrushi2009detecting} explored an estimation-inspection algorithm using logistic regression, and evaluated their design on the testbed of Linux based PLCs, generating a high detection probability with a zero false positive rate.


SCADA systems have the following properties. Firstly, they belong to cyber-physical systems~\cite{rajkumar2010cyber}, which operate real-time with low tolerance on packets delay. Secondly, frequent patching and updating for SCADA intrusion detection modules are unfavourable due to the inflexibility of the infrastructure and the potential negative impact to the whole work process. Thirdly, a high proportion of SCADA devices have limited computing abilities for implementing sophisticated intrusion detection modules. Fourthly, SCADA networks consist of supervisory and control subnetworks. Each sub-network has different characteristics. The hybrid nature of SCADA networks leads to some distinguished characteristics. In particular, the features of field network flows are simpler and more stable, making complex IDS unnecessary. Under these considerations, we design a novel, highly scalable hierarchical online intrusion detection system (HOIDS) for SCADA networks based on machine learning algorithms.
HOIDS is uniquely designed to satisfy the real-time requirement in control systems by utilizing an IDS server-client topology where clients distributed at fields perform intrusion detection using the learning principles generated by a central IDS server.
This is in sharp contrast to existing work where IDS is independently implemented at each node in the network and there are no interactions among the IDS modules.
By selecting the effective data features based on information gain or reducing the dimension of the feature set, the implementation of IDS clients can be simplified to accommodate the SCADA devices without significant impact on the detection accuracy. HOIDS is also flexible to adjust the detection principles for clients based on  practical requirements to improve security.

\section{System Design}
This section describes the proposed IDS designed for a large-scale SCADA system under critical real-time requirements. The architecture and operating principles are described in Subsection II-A.
The detailed detection models, including normal-abnormal binary detection and multi-attack detection based on
logistic regression and 
quasi-Newton optimization algorithm are presented in Subsection II-B. In Subsection II-C, we present the principles of information gain based feature selection and principle component analysis based dimension reduction to reduce the feature set and accelerate detection.

\subsection{HOIDS architecture and principle}

\begin{figure*}[th]
\centering
\includegraphics[width=1.0\textwidth]{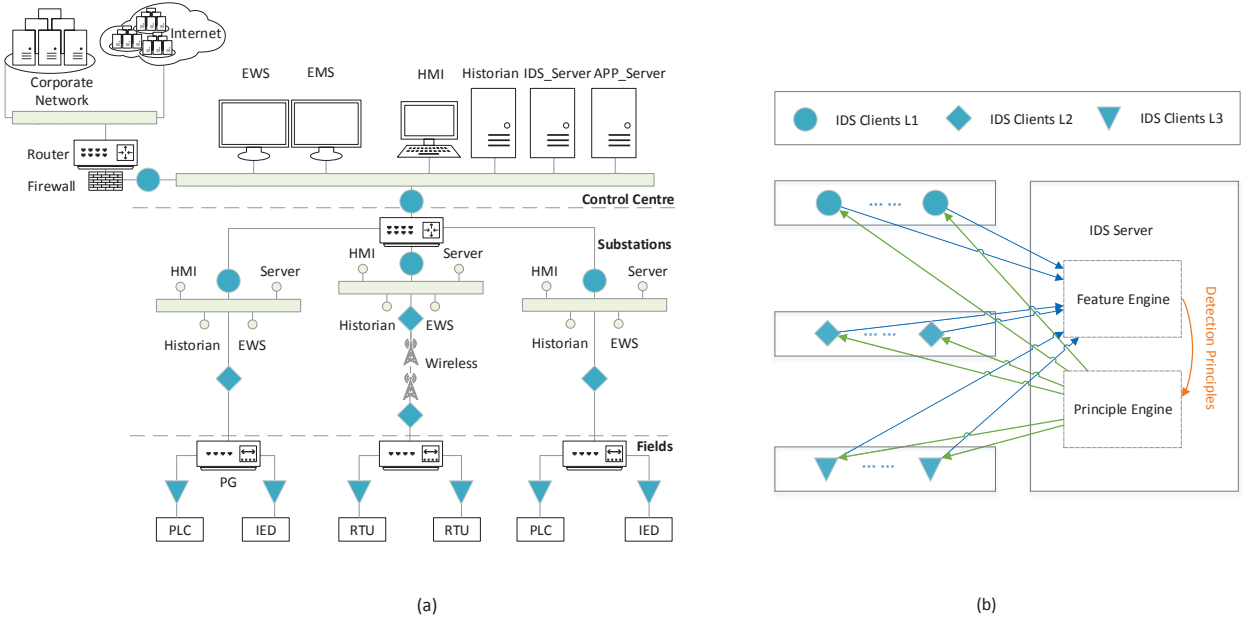}
\caption{(a) HOIDS schemes for SCADA systems. (b) Workflow of the HOIDS data transmission. }\label{fig1}
\end{figure*}

Fig.~\ref{fig1} (a) illustrates a typical SCADA system consisting of the field networks, substations and control centre which is connected to external networks such as the corporate networks and the Internet. In the control centre, there are engineering workstations (EWS), human machine interfaces (HMI), energy management systems (EMS), historians, application servers, etc. The network flow in the control centre is similar to that in the external IT networks. One reason for this similarity is that backend protocols~\cite{knapp2014industrial} such as Open Process Communications (OPC) and Inter-Control Centre Protocol (ICCP) work in a client/server manner supported by TCP/IP over Ethernet. Another reason is that control centre network is usually connected to corporate business networks and the Internet. In the substations, there are EWS, HMI, historians, servers, etc. Field devices consist of intelligent electronic devices (IED), remote terminal units (RTUs), programmable logic controllers (PLC) and many other embedded machines. Generally, the communication protocols used between substations and the field networks are fieldbus protocols. Modbus, DNP3, etc., are widely used, and normally, the networks are isolated from the Internet and other external networks. Protocol gateway (PG) is to translate messages among various protocols.


Due to the hybrid nature of SCADA networks, hierarchical IDS deployment is proposed for different network components as shown in Fig.~\ref{fig1} (a). First, the IDS system is composed of IDS agents residing at all components of the SCADA network. Intrusion detection is realized by machine learning algorithms. Second, a client-server model is adopted with the IDS server at the control centre, and three levels of IDS clients distributed in the control centre, the substation networks and field networks, respectively. The IDS server deployed in the control centre is the command centre of the whole system, and all the IDS clients communicate with and are controlled by the IDS server in the control centre. Such arrangement is motivated by the fact that field devices are usually simple electronics with constrained computing capability, and the amount of data traffic between the field devices and substations is generally limited. Therefore, it will be impossible for these field devices to host IDS agents that execute full machine learning algorithms, unless external computing modules are added specifically for IDS. The client-server model can greatly alleviate the IDS computing needs. It provides flexible IDS complexity for different network components in a SCADA system. Third, the IDS clients can be configured together with signature-based techniques to strengthen security. Furthermore, the IDS server has a holistic view of the whole network and can make adjustment to the detection model of the clients, especially when a certain client detects intrusion, to improve the overall IDS accuracy of all the components in the network.

Fig.~\ref{fig1} (b) illustrates the mechanism of the IDS server in the control centre and the IDS clients distributed at three levels of the network. The IDS server with relatively high computing ability mainly consists of the feature engine and the principle engine. The feature engine is responsible for collecting the features of network flows in different SCADA network components, forming datasets, using the datasets sequentially and periodically to train machine learning algorithms and storing the generated principles for each IDS client to the principle engine. The principle engine is responsible for sending the detection principles  to each IDS client sequentially and periodically.
The IDS clients deployed at different network spots extract features of the real time traffic and apply the received principles to analyze current data flows for intrusion detection, launch alarm if anomaly is detected. The clients also constantly send the extracted features and detection results to the IDS server for periodically updating the training set and detection principles online to ensure the accuracy and timeliness of intrusion detection. The separation of training and detection to server and client save considerable computing resources at the client side. Meanwhile such architecture allows sophisticated machine learning algorithms to be adopted and updated easily for the whole network. Different network elements can use different IDS algorithms, both for training at the server side and detection at the client side. Consequently, this design is effective to reduce the global financial budget for securing large-scale SCADA networks.
We should also note that most devices in the field networks have limited computing abilities and stringent real-time requirement. Therefore, preprocessing features by effective feature selection techniques and dimension reduction methods will accelerate the detection process and achieve online intrusion detection
with minimum impact on the accuracy of intrusion detection (demonstrated in Section~\ref{simul}). Furthermore, when an intrusion is detected at a certain IDS client, the hierarchical design has the potential to coordinate IDS detection at different network elements to enhance SCADA security, for example, by adjusting the choice of feature sets for the IDS client and its adjacent clients.

\subsection{Logistic regression}

In HOIDS design, we apply machine learning techniques to intrusion detection. Specifically, we use logistic regression to classify the training dataset and generate the detection model.
Logistic regression has been a powerful mathematical method for classification problems~\cite{abu2012learning}. 
Generally, different machine learning algorithms have their own advantages. For HOIDS implementation, logistic regression has more advantages over other techniques. For example, the model of logistic regression can be interpreted clearly as a probability, beneficial for result analysis and model adjustment. As opposed to naive Bayes, another probabilistic algorithm that makes classification under the assumption of independent features, logistic regression can generate the classification principle regardless of the correlation among the training features. Compared to SVM and neural networks, the training time of logistic regression is shorter. SVM may generate many supporting vectors in the detection model, reducing detection efficiency if applied in the HOIDS design. Moreover, logistic regression is able to incorporate new training data easily into the current classification model by using stochastic gradient descent method, which is important for industrial applications.
Logistic regression is also efficient to realize multi-classification with low complexity and high accuracy by using the multinomial logistic regression~\cite{bohning1992multinomial} (details in Section~\ref{simulation}), while some other machine learning algorithms rely on the one-against-all approach to achieve multi-classification. Therefore, logistic regression is an appropriate machine learning algorithm that can be applied to industrial network IDS. Next we present the principle of normal-abnormal binary detection and multi-attack detection based on logistic regression.

\subsubsection{Normal-abnormal binary detection}

In normal-abnormal binary detection, the logistic regression classifier identifies  intrusive connections against normal connections. Here, intrusive connections can also be called abnormal connections. Define the feature space of connections as $\mathcal{X} = \mathcal{R}^M$, where $M$ is the number of features for each connection and $\mathcal{R}$ is the set of all real numbers. The classification output space can be expressed as $\mathcal{\gamma} = \{+1, -1\}$, where $+ 1$ is a label representing the abnormal connection and  $-1$ representing the normal connection. The training dataset $\mathcal{D}$ consists of $N$ connections, i.e., $\mathcal{D}=\{\bold{X}, \bold{y}\}=\{(\bold{x}_i, y_i), i = 1, 2, ..., N\}$, where $\bold{X}$ is a ${(M+1)}\times{N}$ matrix representing N connections, and each input connection ${{\bf{x}}_i} = [{x_{i0}}\ \ {x_{i1}} \ \cdot  \cdot  \cdot \  {x_{ij}}  \  \cdot  \cdot  \cdot \  {x_{iM}}]^T$ has ${x}_{i0} = 1$, and $\bold{y}$ is an N-dimensional label vector with each label ${y_i} \in \mathcal{\gamma}$. The classification model weights can be represented by $\bold{w} = {[{w_0} \ \ {w_1} \ \cdot  \cdot  \cdot \  {w_j} \  \cdot  \cdot  \cdot \  {w_M}]^T}$ where $ w_j $ represents the weight of corresponding ${x}_{ij}$. In this case, we need to maximize the conditional probability of getting $\bold{y}$ given the corresponding $\bold{X}$ to generate the classification model, i.e., maximizing
\begin{equation}P({\bf{y}}|{\bf{X}}) = \prod\limits_{i = 1}^N {P({y_i}|{{\bf{x}}_i})}
\label{equa1}\end{equation}
where
\begin{equation}P({y_i}|{{\bf{x}}_i}) = \left\{ \begin{array}{l}
P({y_i} = 1|{{\bf{x}}_i})\ \ \ \ \ \ \ \ \textrm{for}\ {y_i} = + 1\\
1 - P({y_i} = 1|{{\bf{x}}_i})\ \ \ \textrm{for}\ {y_i} =  - 1.
\end{array} \right.\end{equation}

Logistic regression uses the logistic function $\theta (z) = {{{1}} \mathord{\left/
 {\vphantom {{{1}} {(1 + {e^{-z}})}}} \right.
 \kern-\nulldelimiterspace} {(1 + {e^{-z}})}}$
to map the linear combination $z = {\bf{x}}_i^T{\bf{w}}$ to a value between 0 and 1. Let $P({y_i} = 1|{{\bf{x}}_i}) = \theta ({\bf{x}}_i^T{\bf{w}})$, and due to the property of logistic function $\theta ( - z) = 1 - \theta (z)$, we have $P({y_i}|{{\bf{x}}_i}) = \theta ({y_i}{\bf{x}}_i^T{\bf{w}})$. Maximizing the joint conditional probability~\eqref{equa1} is equivalent to minimizing its scaled negative logarithm, therefore we have the minimization of the objective function $F_{bin}$ for normal-abnormal binary classification

\begin{equation}F_{bin}({\bf{w}}) =  -\frac{1}{N} \sum\limits_{i = 1}^N {\ln P({y_i}|{{\bf{x}}_i})}  = \frac{1}{N}\sum\limits_{i = 1}^N {\ln (1 + {e^{ - {y_i}{{\bf{x}}_i^T}{\bf{w}}}})}. \label{EQ_Fbin}\end{equation}

It can be proved easily that the Hessian of \eqref{EQ_Fbin} is positive definite as long as ${\bf{X}} \ne {\bf{0}}$ which holds in all practical cases. Consequently, minimizing \eqref{EQ_Fbin} is a global convex optimization problem and can be solved using conventional gradient-based techniques with a proper line search step.
Here, we adopt the quasi-Newton optimization implemented with the Broyden-Fletcher-Goldfarb-Shanno (BFGS) approach and the back-tracking line search method~\cite{antoniou2007practical}, which has been verified as one of the most efficient optimizers for logistic regression~\cite{minka2003comparison}. The basic quasi-Newton algorithm is as follows: given initial weights ${\bf{w}}_0$ and the tolerance $\epsilon$, the weights will be updated in the next iteration ${{\bf{w}}_{k +1}} = {{\bf{w}}_k} + {{\bm{\delta }}_k}$. And ${{\bm{\delta }}_k} = - {\alpha _k}{{\bf{S}}_k}{{\bf{g}}_k}$ represents the updated step, in which ${\bf{g}}_k$ is the gradient vector of $F_{bin}$ in the ${k^{th}}$ iteration,  ${\alpha_k}$ is a small positive value decreasing the value of $F_{bin}$ in the ${k^{th}}$ iteration, and ${{\bf{S}}_k}$ is an $(M+1) \times (M+1)$ direction matrix. Both ${\alpha_k}$ and ${{\bf{S}}_k}$ can be obtained by different methods. Repeat the iteration until convergence, i.e., ${\lVert{\bm{\delta }}_k}\rVert <  \epsilon$. The BFGS approach is to obtain ${{\bf{S}}_k}$ iteratively according to
\begin{equation}
\begin{array}{rcl}
{{\bf{S}}_0} &=& {\bf{I}}\\
{{\bf{S}}_{k + 1}}& =& {{\bf{S}}_k} + (1 + \frac{{{\bm{\gamma }}_k^T{{\bf{S}}_k}{{\bm{\gamma }}_k}}}{{{\bm{\gamma }}_k^T{{\bm{\delta }}_k}}})\frac{{{{\bm{\delta }}_k}{\bm{\delta }}_k^T}}{{{\bm{\gamma }}_k^T{{\bm{\delta }}_k}}} - \frac{{{{\bm{\delta }}_k}{\bm{\gamma }}_k^T{{\bf{S}}_k} + {{\bf{S}}_k}{{\bm{\gamma }}_k}{\bm{\delta }}_k^T}}{{{\bm{\gamma }}_k^T{{\bm{\delta }}_k}}}
\end{array}
\label{EQ4}
\end{equation}
in which $\bf{I}$ is an identity matrix of size ${(M+1)}$, and ${{\bm{\gamma }}_k} = {{\bf{g}}_{k + 1}} - {{\bf{g}}_k}$. The back-tracking line search is an effective inexact line search method to obtain ${\alpha_k}$ by finding an  ${\alpha}$ satisfying $F_{bin}({{\bf{w}}_{k}} - {\alpha _k}{{\bf{S}}_k}{{\bf{g}}_k}) \leqslant F_{bin}({{\bf{w}}_{k}})$, of which details can be referred to  \cite{boyd2004convex}.



 By minimizing \eqref{EQ_Fbin}, we can obtain the optimal classification weight vector $\bold{w}$ and calculate the probability $P({y_i} = 1|{{\bf{x}}_i})$. A sample will belong to ${y}= {+ 1}$ if the probability exceeds $0.5$, otherwise it will belong to $y=-1$. Since the logistic function is monotonically increasing, compare the value of ${{\bf{x}}_i}{\bf{w}}$ with $0$, and the sample connection will belong to ${y}= {+ 1}$ if the value is positive, otherwise belong to $y=-1$.

\subsubsection{Multi-attack detection}
As mentioned above, logistic regression can realize multi-classification with low complexity and high accuracy by using the multinomial logistic regression
due to the property of the probabilistic model. The representation of the feature space, the number of input data and the number of features are the same as the normal-abnormal binary classification, while the classification output space is $\gamma_{multi} = \{0, 1, ..., K-1\}$, where $K$ represents the number of class types. The classification model weights can be represented by $\bold{W} = \{\bold{w}_0, \bold{w}_1, ..., \bold{w}_{k-2}\}$. Note that $\bold{W}$ is a $(M+1)\times(K-1)$ matrix, which consists of the classification model weights for $(K-1)$ class types. And the objective function of multi-class logistic regression for minimization can be denoted as
\begin{equation}F_{multi}({\bf{W}}) =  - \frac{1}{N} \ln P({\bf{y}}|{\bf{X}}) =  - \frac{1}{N} \sum\limits_{i = 1}^N {\ln P({y_i}|{{\bf{x}}_i})} \label{euq5}\end{equation}
where
\begin{equation}P({y_i}|{{\bf{x}}_i}){\rm{ = }}\left\{ \begin{array}{l}
 {{{e^{{{\bf{x}}_i^T}{{\bf{w}}_{{y_i}}}}}} \mathord{\left/
 {\vphantom {{{e^{{{\bf{x}}_i}{{\bf{w}}_{{y_i}}}}}} {(1 + \sum\limits_{a = 0}^{K - 2} {{e^{{{\bf{x}}_i}{{\bf{w}}_a}}}} )}}} \right.
 \kern-\nulldelimiterspace} {(1 + \sum\limits_{a = 0}^{K - 2} {{e^{{{\bf{x}}_i^T}{{\bf{w}}_a}}}} )}},0 \le {y_i} < K - 1 \\
 {1 \mathord{\left/
 {\vphantom {1 {(1 + \sum\limits_{a = 0}^{K - 2} {{e^{{{\bf{x}}_i}{{\bf{w}}_a}}}} )}}} \right.
 \kern-\nulldelimiterspace} {(1 + \sum\limits_{a = 0}^{K - 2} {{e^{{{\bf{x}}_i^T}{{\bf{w}}_a}}}} )}},{y_i} = K - 1 \\
 \end{array} \right.\label{euq6}\end{equation}
and note that $\sum\limits_{{y_i} = 0}^{K - 1} {P({y_i}|{{\bf{x}}_i})}  = 1$. By minimizing the multi-classification objective function~\eqref{euq5} using the quasi-Newton optimization implemented with the BFGS approach, we can obtain the optimal multi-classification weight matrix $\bold{W}$. To classify a new sample connection, calculate all the $P({y_i}|{{\bf{x}}_i}), y_i = 0, 1, ..., K-1$ based on~\eqref{euq6}, and then the sample connection belongs to the class type with the highest probability. In this way, we can use the generated detection principle to classify the testing dataset. Since the one-against-all approach needs to train the binary classification for $K$ times to generate $K$ model weights for all the classes, multinomial logistic regression is more efficient than the one-against-all approach.

\subsection{Feature selection and dimension reduction}
In HOIDS, especially for the field networks, we apply two methods for preprocessing SCADA network data features.
The first method is to use information gain to denote the significance of each feature, and select features with high information gain to accelerate intrusion detection.
Similar to mutual information, information gain is the reduction in the entropy of labels achieved by partitioning the labels according to a certain feature. For a certain feature, the information from labels will be changed if this feature is not included in the system. As a result, the reduced label entropy is
the information the feature brings to the system.

For the training dataset $\mathcal{D}=\{\bold{X}, \bold{y}\}=\{(\bold{x}_i, y_i), i = 1, 2, ..., N\}$ with $M$ features of the input connections and $K$ classifications of labels, the information entropy $H(\bf{y})$ of the label $\bf{y}$ can be obtained by
 \begin{equation}H({\bf{y}}) =  - \sum\limits_{k = 0}^{K - 1} {P({y_k})} \ln P({y_k})\end{equation}
where $P({y_k}) = {{{N_k}} \mathord{\left/
 {\vphantom {{{N_k}} N}} \right.
 \kern-\nulldelimiterspace} N}$, and ${N_k}$ represents the number of samples of class $k$ in the training dataset.

 Assume a feature ${F_m}$ consisting of values $\{ {f_1},{f_2},...{f_S}\}$, the information gain $IG$ that feature ${F_m}$ brings to the system is
\begin{equation}IG({F_m}) = H({\bf{y}}) - H({\bf{y}}|{F_m})\end{equation}
in which
\begin{equation}H({\bf{y}}|{F_m}) = \sum\limits_{s = 1}^S {P({f_s})H({\bf{y}}|{f_s})} \end{equation}
$P({f_s}) = {{{N_{{f_s}}}} \mathord{\left/
 {\vphantom {{{N_{{f_s}}}} N}} \right.
 \kern-\nulldelimiterspace} N}$, and ${N_{{f_s}}}$ denotes the number of samples which have the value ${f_s}$ in terms of feature ${F_m}$. Usually, a feature with high information gain is preferred over those with low information gain. Here, we exploit the concept of information gain to select a feature subset to accelerate the detection model training and online detection process.

The second method is to use singular value decomposition (SVD) to obtain a low-dimensional approximation of the original feature set \cite{antoniou2007practical}, which is also known as principal component analysis (PCA). We apply PCA to the covariance matrix of the $N$ feature vectors. The covariance matrix can be calculated as

\begin{equation}{\bf{C}} = \frac{1}{{N - 1}}\sum\limits_{i = 1}^N {({{\bf{x}}_i} - {\bf{\bar x}}){{({{\bf{x}}_i} - {\bf{\bar x}})}^T}}\end{equation}
where ${\bf{\bar x}} = \frac{1}{N}\sum\limits_{i = 1}^N {{{\bf{x}}_i}}$ denotes the average vector of the $N$ feature vectors. Since ${\bf{C}}$ is at least positive semidefinite, its SVD is the same as its eigen decomposition. We can obtain an approximation of variance matrix ${\bf{C}}$ by considering only the $K$ largest eigenvalues and the corresponding eigenvectors as
\begin{equation}{\bf{C}} = {\bf{U}}{\bf{S}}{{\bf{U}}^T} \approx {{\bf{U}}_K}{{\bf{S}}_K}{\bf{U}}_K^T\end{equation}
in which ${\bf{U}} = [{{\bf{u}}_1}\ {{\bf{u}}_2} \cdot  \cdot  \cdot {{\bf{u}}_M}]$ is an orthogonal matrix, ${\bf{S}} = diag\{ {\sigma _1},{\sigma _2}, \cdot  \cdot  \cdot ,{\sigma _M}\}$ with non-negative eigenvalues in a descending order, ${{\bf{U}}_K}$ contains the first $K$ vectors of ${\bf{U}}$, and ${{\bf{S}}_K}$ is the diagonal matrix containing the first $K$ largest eigenvalues. Compute the projection of the variation $({{\bf{x}}_i} - {\bf{\bar x}})$ onto the K-dimensional subspace spanned by ${{\bf{U}}_K}$ as
\begin{equation}{{\bf{z}}_i} = {\bf{U}}_K^T({{\bf{x}}_i} - {\bf{\bar x}}).\end{equation} 
We use ${{\bf{z}}_i}$ as the new features for each connection. Since ${{\bf{U}}_K}$ includes $K$ orthogonal vectors, the new features generated by PCA are uncorrelated. The number of principle components $K$ can be chosen by finding the smallest $K$ satisfying
${{{\sum\limits_{i = 1}^K {\sigma _i}}} \mathord{\left/
 {\vphantom {{{\sum\limits_{i = 1}^K {\sigma _i}}} {\sum\limits_{i = 1}^M {\sigma _i}}}} \right.
 \kern-\nulldelimiterspace} {\sum\limits_{i = 1}^M {\sigma _i}}} \ge \rho$, where $\rho$ is the rate of variance retained. This method is specifically effective when the original features
 are heterogeneous 
 and some may be highly correlated.

For the detection with a certain feature set, we use recall and precision to measure the performance, which are the critical performance measures for IDS. For the IDS binary classification, true positive (TP), false negative (FN), false positive (FP) and true negative (TN) denote the quantities of intrusions identified as intrusions, intrusions identified as normal, normal connections identified as intrusions and normal connections identified as normal, respectively. Note that for multi-classification, we define TP, FN, FP and TN as the quantities of intrusions identified as correct intrusion types, intrusions identified as normal connections, normal connections identified as intrusions or intrusions incorrectly identified as different intrusion types and normal connections identified as normal, respectively. Recall ($r$) is defined as $TP / (TP + FN)$, and precision ($p$) is defined as $TP / (TP + FP)$. For IDS, $r$ is important, while $p$ can not be ignored as well, since intrusions should be detected as many as possible, while alarms are supposed to be real intrusions as many as possible. Specifically, we use $10 \times 10$-fold cross-validation (CV) to evaluate $r$ and $p$ for each selected feature combination. Compared with $10$-fold CV, the most widely used validation procedure, $10 \times 10$-fold CV can obtain more reliable performance estimation since more estimates are always preferred \cite{bouckaert2003choosing}.
Estimating the mean of $r$ for $10 \times 10$-fold CV are compared using confidence intervals defined by
$\bar r \pm {t_{\frac{\alpha }{2}}}(n - 1)\frac{s}{{\sqrt n }}$,
in which $\bar r$ and $s$ are the mean and standard deviation of $r$ of the CV samples, ${t_{\frac{\alpha }{2}}}(n - 1)$ is the value of $t$ distribution at $(n-1)$ degrees of freedom for a $(1 - \alpha)$ confidence interval, and n is the size of CV samples. Estimation for the mean of $p$ is the same as that for $r$.
%
%
In this way, we can choose the model of preprocessing the original features based on the performance estimation of CV.
\section{Numerical results\label{simul}}
This section presents the classification results based on the logistic regression algorithm to simulate the intrusion detection for SCADA networks. Note that network data flows have unique characteristics specific to that type of network. For example, the network data flows in a control centre are similar to traditional IT networks, and thus the corresponding features can be extracted by the corresponding IDS clients in a similar way to those in IT networks. Therefore, we simulate the network data in the control centre by exploiting the KDD99 dataset~\cite{kdd}, the most widely used dataset for evaluating the performance of an IDS designed for IT networks. On the other hand, the network traffic in the substation and field network can be simulated by the ICS dataset~\cite{ICSDataset},~\cite{morris2011testbed}.
All the experiments are implemented by Matlab and run on a server with Intel Xeon 8-core processor E5-2670 and 64 GB RAM.

\subsection{Simulation of the SCADA control centre network\label{simulation}}
In this section, we present the simulation results of a SCADA control centre network using the KDD99 dataset. In our experiment, we exploit the KDD99 (10 percent) training dataset for training and KDD99 testing dataset for testing. The training dataset consists of $494,021$ samples with $41$ features ($3$ nominal and $38$ numerical) and $22$ different types of attacks (e.g., Back, Land, Neptune and Smurf) that fall into four categories: Denial of service (DOS), Probe, Remote to local (R2L) and User to root (U2R). In the training dataset, there are $97,278$ ($19.69\%$) normal connections, $391,458$ ($79.24\%$) DOS, $4,107$ ($0.83\%$) Probe, $1,126$ ($0.23\%$) R2L and $52$ ($0.01\%$) U2R connections. The testing dataset (consisting of $311,029$ samples with $41$ features) contains $22$ attack types existed in the training dataset and additional $17$ different kinds of attacks. In the testing dataset, there are $60,593$ ($19.48\%$) normal connections, $229,853$ ($73.90\%$) DOS, $4,166$ ($1.34\%$) Probe, $16,189$ ($5.20\%$) R2L and $228$ ($0.07\%$) U2R connections. The fact that the testing dataset does not have the same probability distribution as the training dataset, to some extent, makes the detection process close-to-realistic scenarios. All the 41 features, including time- and host-based traffic features, are derived from the characteristics of the network data flow. In our simulation, we exploit $38$ numerical features for training and testing.
\begin{table}[h]
\caption{Classification for the KDD99 testing dataset}
\label{table_kdd}
\begin{center}
\begin{tabular}
{ |K{0.02cm} | K{0.50cm} | K{0.65cm} | K{0.76cm} | K{0.56cm} | K{0.56cm} | K{0.56cm} | K{0.58cm} | K{0.58cm} |}
\cline{3-7}
\multicolumn{2}{c |}{} & \multicolumn{5}{c |}{Predicted Class} & \multicolumn{2}{c}{ }  \\
\cline{3-9}
\multicolumn{2}{c|}{}  & Norm & DOS &  Probe &  R2L  &  U2R & $r$ & $r_{PN}$
\\ \hline \multirow{5}{0.02cm}{\begin{sideways} Actual Class~ \end{sideways}}& Norm & 59579  & 790 & 172 & 44 & 8 & 0.983 & 0.995
\\ 
\cline{2-9}
& DOS & 6380  & 223451  &  20  &  0  &  2 & 0.972 & 0.969
\\ 
\cline{2-9}
& Probe & 1064  & 96 & 3004 & 0 & 2 & 0.721 & 0.730
\\ 
\cline{2-9}
& R2L & 16137  & 4  &  18  &  25  & 5 & 0.002 & 0.107
\\ 
\cline{2-9}
& U2R & 197  & 4 & 0 & 6 & 21 & 0.092 & 0.066
\\ \hline
\multicolumn{1}{c|}{} & $p$ & 0.715  & 0.996 & 0.935 & 0.333 & 0.553 & &\multicolumn{1}{c |}{}
\\
\cline{2-9}
\multicolumn{1}{c|}{} & $p_{PN}$ & 0.730  & 0.9995 & 0.925 & 0.880 & 0.105 & &\multicolumn{1}{c |}{}
\\ 
\cline{2-9}
\end{tabular}
\end{center}
\end{table}
Using the multinomial logistic regression to classify the training dataset
and applying the generated detection principle to the testing dataset, we get the confusion matrix in Table~\ref{table_kdd}.
Compare these results (recall (r) and precision (p)) with those obtained by the PNrule method~\cite{agarwal2000pnrule} (recall (r$_{PN}$) and precision (p$_{PN}$)),
a rule-based classifier applicable to scenarios where different classes have very different distributions in training data. 
As shown, the overall performance of both methods are consistent. This confirms the validity of multinomial logistic regression. Logistic regression shows higher recall and precision for U2R than PNrule, while PNrule achieves higher recall and precision for R2L. In terms of R2L, logistic regression may benefit from more training samples, since the percentage of all R2L attacks in the training dataset is only $0.23\%$, while $5.20\%$ in the testing dataset with additional $7$ attack types not shown in the training dataset.


\begin{figure}
\centering
\includegraphics[width=3.7in]{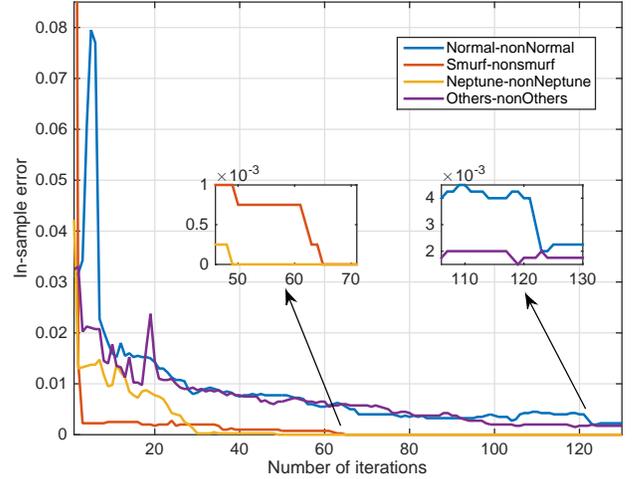}
\caption{One-against-all for KDD99 (Sampled)}
\label{fig_one_all}
\end{figure}

\begin{figure}
\begin{center}
\includegraphics[width=3.7in]{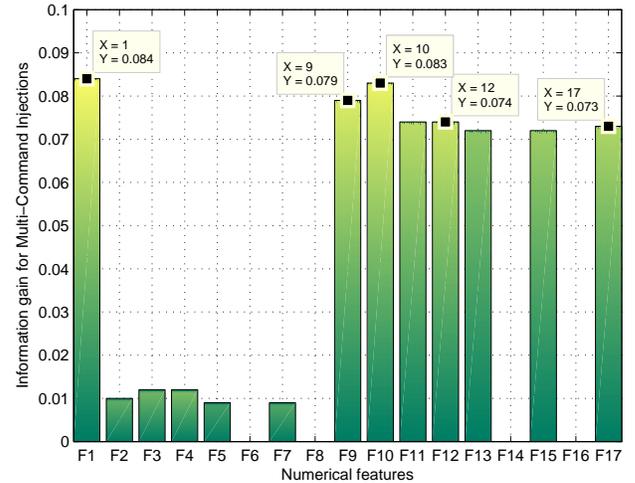}
\caption{Information Gain for Multi-Command Injections (Entropy = 0.084)}
\label{histogram_command_features}
\end{center}
\end{figure}
Next, we verify the performance of multinomial logistic regression by comparing it with the one-against-all method. To simplify the simulation, we randomly sample the training dataset. Since the percentages of attack categories of DOS ($79.24\%$), Probe ($0.83\%$), R2L ($0.23\%$) and U2R ($0.01\%$) are uneven, we classify the attacks into $4$ categories as Normal ($97,278$), Smurf ($280,790$), Neptune ($107,201$) and other attacks ($8,752$), which has the percentage of Normal ($19.69\%$), Smurf ($56.84\%$), Neptune ($21.70\%$) and Others ($1.77\%$). With the same percentage as the KDD99 (10 percent) dataset, we randomly choose $800$ Normal, $2,320$ Smurf, $800$ Neptune and 80 Others to form the new training dataset ($4,000\times38$), and $1,000$ Normal, $2,900$ Smurf, $1,000$ Neptune and $100$ Others to form the new testing dataset ($5,000\times38$) such that the probability distribution of the testing dataset become similar to that of the training dataset.
We first use logistic regression one-against-all binary classification to realize the multi-classification. The in-sample error ($E_{in}$, the ratio of misclassified samples to the total samples in the training dataset) changing with the number of iterations for the new training dataset is shown in Fig.~\ref{fig_one_all}. After $130$ iterations, $E_{in}$s are $0$, $0$, $0.001$, $0.002$ for normal-nonnormal, smurf-nonsmurf, neptune-nonneptune and others-nonothers binary classifications, respectively. The global $E_{in}$ for the training dataset is $0.002$.
Out of $4,000$ training samples, $3,993$ samples are classified correctly.
The out-of-sample error ($E_{out}$, the ratio of misclassified samples to the total samples in the testing dataset) for the new testing dataset is $0.005$.
Out of $5,000$ testing samples, $4,974$ samples are classified correctly.
Next, we use the multinomial logistic regression method to achieve the multi-classification. The global $E_{in}$ for the training dataset is $0$. And $E_{out}$ for the testing dataset is $0.0046$.
Out of $5,000$ testing samples, $4,977$ samples are classified correctly.
The multinomial logistic regression
outperforms in terms of efficiency and accuracy.

\subsection{Simulation of SCADA substation and field networks}
To test the SCADA substation and field networks, we exploit the ICS dataset gathered from a gas pipeline system
~\cite{{morris2011testbed}} as mentioned before. This dataset has $26$ features, including $17$ numerical features such as Invalid Function Code (a binary bit indicating the validity of function code), Pump State (a binary bit indicating the state of the pump: on or off) and so on. Here we take the Multi-class Command Injection dataset, an important part of the above
ICS dataset, as an example to analyze the ICS dataset.
The Multi-class Command Injection dataset models the issued commands from the master to control the gas pipeline system,
consisting of $28,086$ Good commands, $2$ Address Scan attacks, $9$ Function Code Scan attacks, $198$ Illegal Setpoint attacks and $49$ PID Modification attacks.

To study the impact of features for the classifier, information gains of each numerical feature are shown in Fig.~\ref{histogram_command_features}. After further analyzing the dataset, we find that F6 PID Rate, F8 Pipeline PSI, F14 delta PID Rate and F16 delta Pipeline PSI are all constant, F3 PID Cycle Time, F4 PID Deadband, F5 PID Gain, F7 PID Reset are highly correlated, and F11 delta PID Cycle Time, F12 delta PID Deadband, F13 delta PID Gain, F15 delta PID Reset are highly correlated. Therefore, 7 features are left in the dataset for analysis after removing the constant features and highly correlated features.

\begin{figure}
\begin{center}
\includegraphics[width=3.7in]{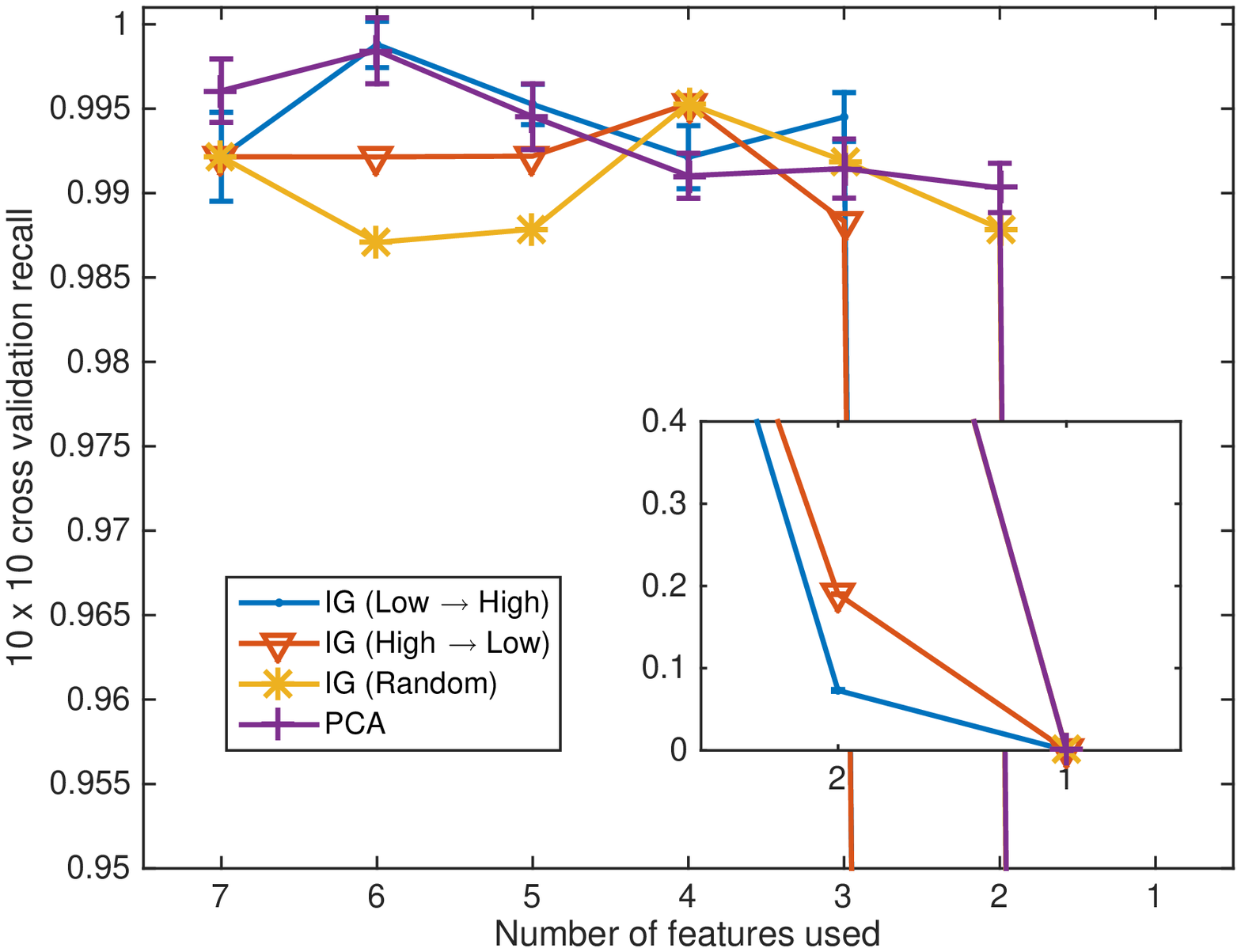}
\caption{Cross validation performance for Multi-Command Injections}
\label{recall_precision performance1}
\end{center}
\end{figure}

\begin{figure}
\begin{center}
\includegraphics[width=3.7in]{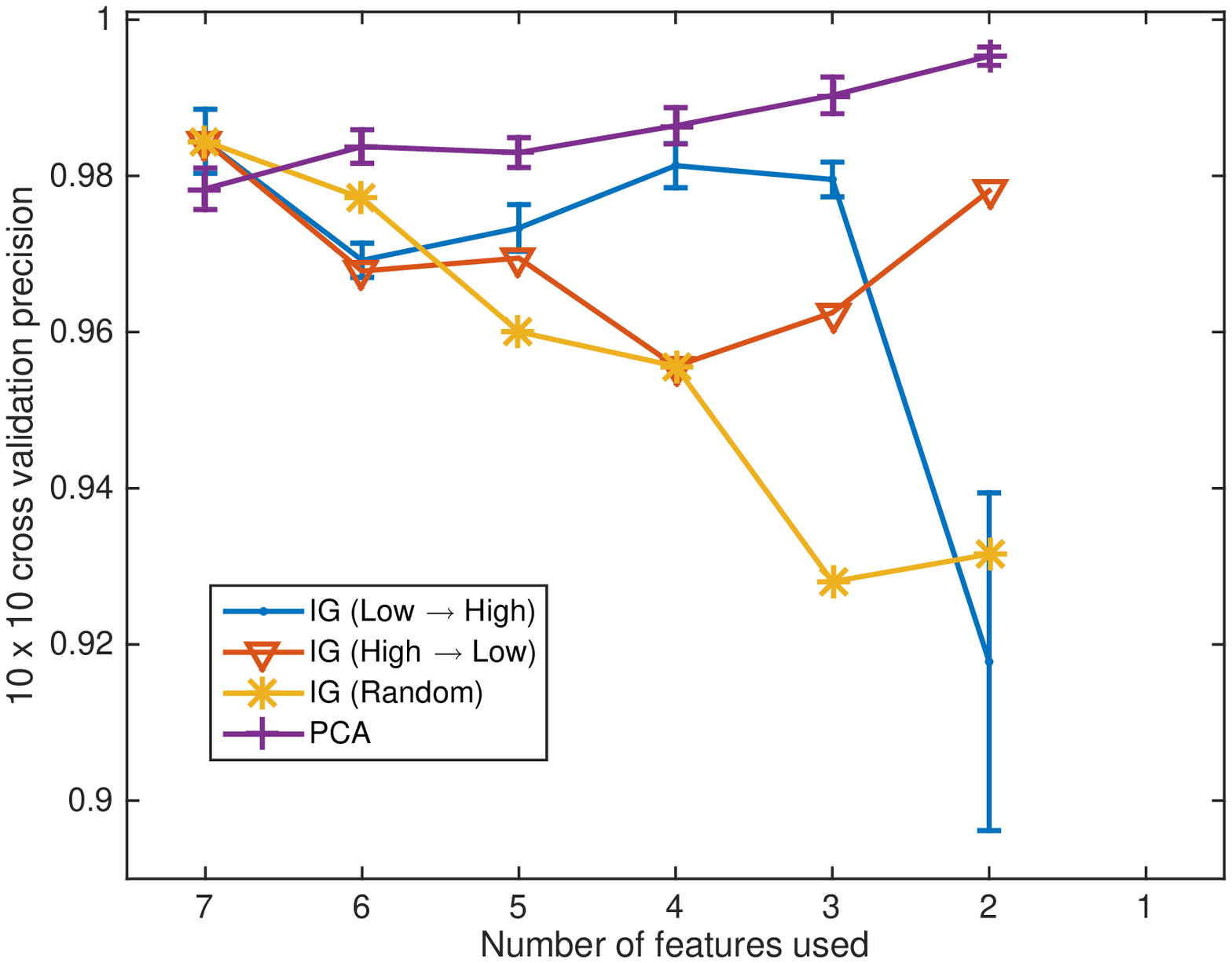}
\caption{Cross validation performance for Multi-Command Injections}
\label{recall_precision performance2}
\end{center}
\end{figure}

We present the $10 \times 10$-fold CV performance for the dataset. The recall and precision for $10 \times 10$-fold CV are presented using $95\%$ confidence intervals in Fig.~\ref{recall_precision performance1} and Fig.~\ref{recall_precision performance2}, respectively.
The features are reduced one by one in three different ways: from the features with lower $IG$ to higher $IG$ (blue lines), from the features with higher $IG$ to lower $IG$ (red lines) and from a random order (yellow lines). The $95\%$ confidence intervals of the last two reducing ways are all less than $0.003$,
not presented on the figures for clearer presentation. In Fig.~\ref{recall_precision performance2}, precisions have no meaning when all the connections are classified as good connections, because $TP + FP$ is equal to $0$, referring to the definition of precision.
From Fig.~\ref{recall_precision performance1} and Fig.~\ref{recall_precision performance2},
we can see that reducing features randomly or from those with higher $IG$ can make both recall and precision much
worse than the original feature set, while reducing features with lower $IG$ can keep good performance even when only 4 or 3 features are left in the feature set. Note that although the recalls of red and yellow lines when 4 features are left are marginally
above the blue line with overlapping confidence intervals, their precisions are notably
less than the blue line. Therefore, we can conclude that reducing the features with lower $IG$ while keeping a few features with higher $IG$ is an effective way to select feature subsets, able to accelerate training and detection while keeping high recalls and precisions.
Dimension reduction based on PCA is also applied to the ICS dataset, shown as the purple lines in Fig.~\ref{recall_precision performance1} and Fig.~\ref{recall_precision performance2}. The number of principle components is reduced from $7$, since the first $7$ eigenvalues keep about $100\%$ variance. The order of reducing new features is from the ones with lower eigenvalues. The $10 \times 10$-fold CV performance is obtained by using the new features generated by projection. We can see that when the number of new features is $6$, the $10 \times 10$-fold CV performance is better than reducing the original features with lower $IG$. Even when the number of new features is reduced to $2$, the recall does not show a significant decline while having an increased precision.
Therefore, for the control networks, we can use reduced feature sets by PCA or IG criterion to accelerate and simplify the IDS process without much degradation of the detection performance.


\begin{figure}
\begin{center}
\includegraphics[width=3.7in]{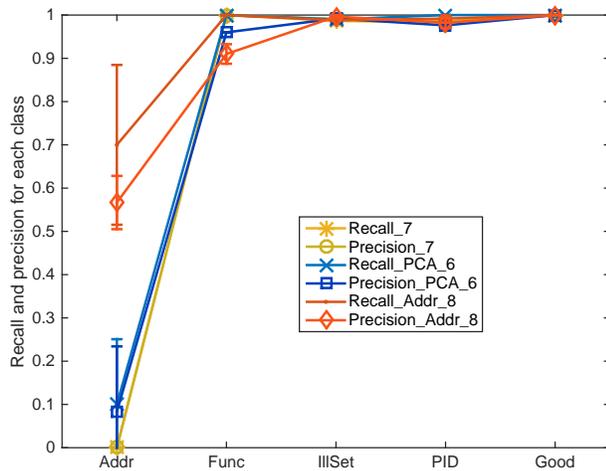}
\caption{Performance of Each Class for Multi-Command Injections}
\label{recall_precision performance3}
\end{center}
\end{figure}

Furthermore, recalls and precisions for each class are presented in Fig.~\ref{recall_precision performance3}. The yellow lines are the results of using the 7 features, comparable with the best work among $6$ machine learning algorithms in~\cite{{beaver2013evaluation}}, which used $12$ features in the dataset and 2 of them are exploited in our dataset. The blue lines are obtained by using 6 new features generated by PCA, which improves the recall and precision of detecting the Address Scan attack to around $0.1$.
Although the number of this attack is only $2$ in the dataset, it is still meaningful to improve the performance of detecting them.
Moreover, we process the Address feature among nominal features and
add it into our preprocessed dataset with 7 independent features, making a total of 8 features. The results are presented with the red lines. We can see that the recall and precision of detecting the Address Scan attack get improved significantly to 0.700 and 0.567, respectively.


\section{Conclusion}

This paper has proposed a novel hierarchical online IDS based on logistic regression and quasi-Newton optimization algorithm. The design of the IDS server in the control centre and IDS clients distributed in the control centre, substation and field networks can secure all cyber assets within the SCADA systems intelligently, while at the same time reducing the financial budget for the large scale systems. Furthermore, the design of intrusion detection with smaller sets of features based on feature selection and dimension reduction can accelerate the detection process to satisfy the real-time requirements of SCADA systems and facilitate the implementation of the hardware devices.

In the future work, we will implement HOIDS  on a practical SCADA testbed to record the network data flows, extract significant features, generate detection rules based on machine learning algorithms and realize online detection.


\end{document}